\documentclass[aip,jmp,12pt,superscriptaddress]{revtex4-1}
\usepackage{graphicx}
\usepackage{amsmath}
\usepackage{amsfonts}
\usepackage{bbm}
\usepackage{hyperref}

\begin{document}

\title{A generalization of the Entropy Power Inequality to Bosonic Quantum Systems}
\author{G. De Palma}
\affiliation{NEST, Scuola Normale Superiore and Istituto Nanoscienze-CNR, I-56127 Pisa,
Italy.}
\affiliation{INFN, Pisa, Italy}
\author{ A. Mari}
\affiliation{NEST, Scuola Normale Superiore and Istituto Nanoscienze-CNR, I-56127 Pisa,
Italy.}
\author{V. Giovannetti}
\affiliation{NEST, Scuola Normale Superiore and Istituto Nanoscienze-CNR, I-56127 Pisa,
Italy.}

\begin{abstract}
In most communication schemes information is transmitted via travelling modes of electromagnetic radiation. These modes are unavoidably subject to environmental noise along any physical transmission medium and the quality of the communication channel strongly depends on the minimum noise achievable at the output. For classical signals such noise can be rigorously quantified in terms of the associated Shannon entropy and it is subject to a fundamental lower bound called entropy power inequality. Electromagnetic fields are however quantum mechanical systems and then, especially in low intensity signals, the quantum nature of the information carrier cannot be neglected and many important  results derived within classical information theory require non-trivial extensions to the quantum regime. Here we prove one possible generalization of the Entropy Power Inequality to quantum bosonic systems. The impact of this inequality in quantum information theory is potentially large and some relevant implications are considered in this work.
\end{abstract}

\maketitle
\section{Introduction}
In standard communication schemes, even if based on a digital encoding, the signals which are physically transmitted are intrinsically analogical in the sense that they can assume a continuous set of values. For example, the usual paradigm
is the transmission of information via amplitude and phase modulation of an electromagnetic field.
In general, a continuous signal with $k$ components
can be modeled by a random variable $X$ with values in $\mathbb R^k$ associated with a probability measure $d\mu(x)=p(x)d^kx$ on $\mathbb R^k$.
For example, a single mode of electromagnetic radiation is determined by a complex amplitude and therefore it can be classically described by a random variable $X$ with $k=2$ real
components.
The Shannon differential entropy~\cite{Shannon,dembo} of a general random variable $X$ is defined as
\begin{equation}
H(X)=- \int_{\mathbb R^k} p(x) \ln p(x)\; d^k x\;, \quad x \in \mathbb R^k\;,
\end{equation}
and plays a fundamental role in information theory. Indeed depending on the context $H(X)$ quantifies the
noise affecting the signal or,
alternatively, the amount of information potentially encoded in the variable $X$.

\begin{figure}[t]
\includegraphics[width=0.8 \columnwidth]{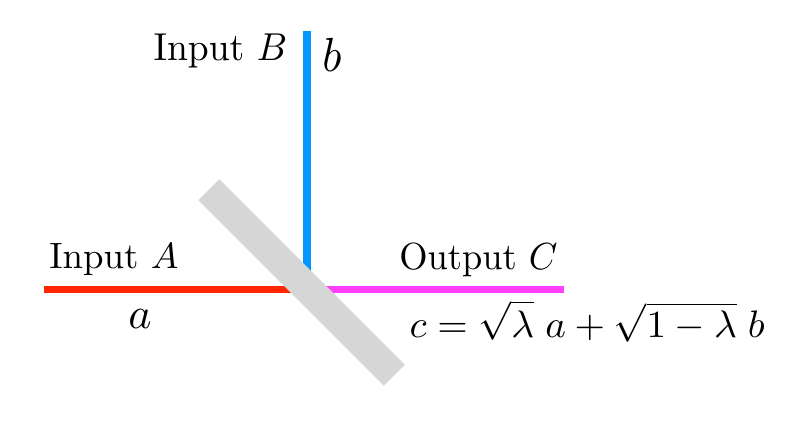}
\caption{Graphical representation of the coherent mixing of the two inputs $A$ and $B$. For the quantum mechanical analogue the two input signals correspond to
 electromagnetic modes which  are coherently  mixed at a beam--splitter
of transmissivity $\lambda$. The entropy of the output signal is lower bounded by a function of the input entropies via the quantum entropy power inequality defined in Eq.~(\ref{qEPI}). } \label{mixing}
\end{figure}

Now, assume to {\it mix} two random variables $A$ and $B$ and to get the new variable $C=\sqrt{\lambda}\;A + \sqrt{1- \lambda}\;B$ with  $\lambda \in [0,1]$ (see Fig.\ \ref{mixing}).
For example this is exactly the situation
in which two optical signals are physically mixed via a beam--splitter of transmissivity $\lambda$. What can be said about the entropy of the output variable $C$?
It can be shown that, if the inputs $A$ and $B$ are independent, the following {\it Entropy Power Inequality} (EPI) holds \cite{stam,blachman}
 \begin{equation} \label{cEPI}
e^{2 H(C)/k}\ge \lambda\; e^{2 H(A)/k}+ (1- \lambda)\;e^{2 H(B)/k}\;,
\end{equation}
stating that for fixed $H(A)$, $H(B)$, the output entropy $H(C)$ is minimized taking $A$ and $B$ Gaussian with proportional covariance matrices.
This is basically a lower bound on $H(C)$ and the name {\it entropy power} is motivated by the fact that if $p(x)$ is a product of $k$ equal isotropic Gaussians one has $\frac{1}{2 \pi e}e^{2 H(X)/k}=\sigma^2$, where
$\sigma^2$ is the variance of each Gaussian which is usually identified with the energy or {\it power} of the signal~\cite{Shannon}.
In the context of (classical) probability theory, several equivalent reformulations~\cite{dembo} and
generalizations~\cite{verdu,rioul,guo}  of Eq.~(\ref{cEPI})  have been proposed,
whose proofs have recently renewed the interest in the field.
As a matter of fact,  these inequalities play a fundamental role in classical information theory, by providing computable bounds for the information capacities of various models of noisy channels~\cite{Shannon,bergmans,LYC}.

The need for a quantum version of the EPI has arisen in the attempt of solving some fundamental problems in quantum communication theory. In particular the EPI has come into play when it has been realized that a suitable generalization to the quantum setting, called \emph{Entropy Photon number Inequality} (EPnI) \cite{EPnIguha,guha}, would directly imply the solution of several optimization problems, including the determination of the classical capacity of Gaussian channels and of the capacity region of the bosonic broadcast channel \cite{guhaieee,guhabroadcast}.
Up to now the EPnI is still unproved and, while the classical capacity has been recently computed \cite{ghgp,ggpch} by proving the bosonic  minimum output entropy conjecture \cite{CONJ1}, the exact capacity region of  the broadcast channel remains undetermined. In 2012 another quantum generalization of the EPI has been proposed, called \emph{quantum Entropy Power Inequality} (qEPI) \cite{ks,ks-natphot}, together with its proof valid only for the $50:50$ beam--splitter corresponding to the case $\lambda=\frac{1}{2}$. The contribution of this paper is to show the validity of this inequality for any beam--splitter and to extend it also to the quantum amplifier.

The qEPI proved in this work, while directly giving tight bounds on several entropic quantities, also constitutes a
potentially powerful tool which could  be used in quantum information theory in the same spirit
in which the classical EPI was instrumental in deriving important classical results like: a bound to the capacity of non--Gaussian channels \cite{Shannon}, the convergence of the central limit theorem \cite{CLT},  the secrecy capacity of the Gaussian wire--tap channel \cite{LYC}, the capacity region of broadcast channels \cite{bergmans}, \emph{etc.}.  In this work we consider some of the direct consequences of the qEPI and we hope to stimulate the research of other important implications in the field.

\section{The quantum Entropy Power Inequality and its proof}
In order to define the quantum mechanical analogue of Eq.\ (\ref{cEPI}), we follow the reasoning line of Ref.s \ \onlinecite{ks,EPnIguha,guha,ks-natphot} where the classical random variable $X$ is replaced by a collection of independent bosonic modes. Specifically consider $n$ optical modes described by
$a_1, a_2, ..., a_n$ annihilation operators obeying the bosonic commutation rules $\left[a_i,\,a_j^\dag\right]=\delta_{ij}$ \cite{BRAU, WEEDB}. This system represents the quantum analogue of the classical random variable $A$. We observe that, since the phase space of each mode is 2-dimensional, the total number of phase space variables is $2n$ and this should be identified with the number $k$ appearing in the classical EPI\ (\ref{cEPI}). A similar collection of bosonic modes $b_1, b_2, ..., b_n$ will play the role of system $B$.
The natural way of mixing the two signals is via a beam--splitter of transmissivity $\lambda$~\cite{WALLS}, which in the quantum optics formalism is represented by the unitary operation $U=e^{\arctan\sqrt{\frac{1-\lambda}{\lambda}} \sum_j \left(a_j^\dag b_j - a_j b_j^\dag\right)}$. This produces $n$ output modes with bosonic operators
\begin{equation}
c_j=\sqrt{\lambda}\;a_j + \sqrt{1- \lambda}\;b_j, \quad j=1,2, ...,n\;.
\end{equation}
In the Schr\"odinger picture the above transformation corresponds to a quantum channel~\cite{channels} mapping the input state $\rho_{AB}$ to the output state
\begin{equation}\label{U}
\rho_C= \mathcal E(\rho_{AB})= \mathrm{Tr}_B\left[ U \rho_{AB} U^\dag\right]\;,
\end{equation}
where the partial trace $\mathrm{Tr}_B$ stems for the fact that we discard one of the two output ports of the beam--splitter.
We consider the case of independent inputs $A$, $B$, with a factorized density matrix $\rho_{AB}=\rho_{A}\otimes\rho_B$. The {\it quantum Entropy Power Inequality} (qEPI) reads then
\begin{equation} \label{qEPI}
e^{S(\rho_C)/n}\ge \lambda\; e^{S(\rho_A)/n}+ (1- \lambda)\;e^{S(\rho_B)/n}\;,
\end{equation}
where the classical Shannon entropy has been replaced by the quantum von Neumann entropy $S(\rho)= -\mathrm{Tr} \left[ \rho \ln \rho \right]$.
Unlike the classical case, the qEPI is \emph{not} saturated by Gaussian states with proportional covariance matrices, unless they have the same entropy. The qEPI (\ref{qEPI}) was conjectured in Ref.\ \onlinecite{ks-natphot} where it was shown to hold only for the special case of $\lambda=\frac{1}{2}$. In this work we prove that inequality (\ref{qEPI}) is indeed
valid for every $\lambda$. Moreover we extend the qEPI to the case in which the two input states are mixed via a {\it quantum amplifier}, {\it i.e.}\ when the unitary $U$ is
replaced by the two-mode squeezing~\cite{WALLS} operation $U'=e^{\mathrm{arctanh}\sqrt{\frac{\kappa-1}{\kappa}} \sum_j \left(a_j^\dag b_j^\dag -a_j b_j\right)}$ with $\kappa \in [1,\infty]$. In this case the modes $a_j$ are
amplified and the modes $b_j$ are phase-conjugated. In the Heisenberg picture we get
\begin{equation}
c_j=\sqrt{\kappa} \;a_j + \sqrt{\kappa-1}\; b_j^\dag, \quad j=1,2, ...,n\;,
\end{equation}
and the amplifier version of the qEPI becomes
\begin{equation} \label{qEPIk}
e^{S(\rho_C)/n}\ge \kappa\; e^{S(\rho_A)/n}+ (\kappa-1)\;e^{S(\rho_B)/n}\;.
\end{equation}

In the rest of the paper we are going to prove the validity of both inequalities~(\ref{qEPI}) and ~(\ref{qEPIk}) and to show some of their direct implications.

\subsection{Properties of quantum Fisher information}
Almost all classical proofs \cite{rioul} of the EPI are based on two properties of the Fisher information: the {\it Fisher information inequality} (or {\it Stam inequality} \cite{blachman, STAMINEQ}) and the {\it de Bruijn identity} \cite{stam}.
 Here we follow the approach of \onlinecite{ks} in order to generalize such properties to quantum systems. Given a smooth family of states $\theta\mapsto \rho^{(\theta)}$ the
 associated quantum Fisher information can be defined in terms of the relative entropy:
 \begin{equation}
\left.J\left(\rho^{(\theta)};\theta\right)\right|_{\theta=0}\equiv \left.\frac{d^2}{d\theta^2}S\left(\left.\rho^{(0)}\right\|\rho^{(\theta)}\right)\right|_{\theta=0}\;,
\end{equation}
 where $S(\rho_1 || \rho_2)\equiv \textrm{Tr}\left[\rho_1 \left(\ln \rho_1-\ln \rho_2\right)\right]$. Since the relative entropy is non-negative and vanishes for $\theta=0$,
we necessarily have $J\left(\rho^{(\theta)};\theta\right) \ge 0$ and from the definition it is clear that $J\left(\rho^{(c \theta)};\theta\right) =c^2 J\left(\rho^{(\theta)};\theta\right)$. Moreover, from the data processing inequality for the relative entropy its counterpart for the quantum Fisher information follows:
for every quantum channel $\mathcal E$, one has $J\left(\mathcal E \left(\rho^{(\theta)}\right);\theta\right) \le J\left(\rho^{(\theta)};\theta\right)$  \cite{ks}.
For our purposes, the relevant cases are when $\theta$ is associated with translations along the phase space axes, {\it i.e.}  $\rho^{\left(q,P_j\right)}= e^{ i q P_j} \rho\ e^{-i q P_j} $ and
$\rho^{\left(p,Q_j\right)}= e^{- i p Q_j} \rho\  e^{i p Q_j} $, where as usual $Q_j \equiv\left.\left(a_j+a_j^\dag\right)\right/\sqrt{2}$, $P_j \equiv i \left.\left(a_j^\dag -a_j\right)\right/\sqrt{2}$. In this situation one can generalize important results form classical information theory.
In particular if two input states are mixed via a beam--splitter or via a quantum amplifier as described in Eq.\ (\ref{U}), one can derive from the data processing inequality the quantum version of the
 {\it Stam inequality}:
\begin{equation}\label{FII}
\frac{1}{J_C} \ge  \frac{\lambda _A }{J_A}  + \frac{\lambda _B }{J_B}\;,
\end{equation}
where $J=\sum_j J\left(\rho^{\left(p,Q_j\right)};p\right)+J\left(\rho^{\left(q,P_j\right)};q\right)$ and
\begin{align}
\lambda_A\equiv \lambda \qquad &\lambda_B \equiv 1-\lambda\qquad&\text{(beam--splitter)}\\
\lambda_A\equiv \kappa \qquad &\lambda_B \equiv \kappa-1\qquad&\text{(amplifier)}\;.
\end{align}
The proof of \eqref{FII} for the beam--splitter in the special case $\lambda=\frac{1}{2}$ was given in Ref. \onlinecite{ks}. The key point of this paper is the generalization of this proof to any beam--splitter and amplifier, crucial for
the derivation of the qEPIs \eqref{qEPI} and \eqref{qEPIk}. In \onlinecite{ks}, \eqref{FII} is derived from the inequality
\begin{eqnarray}
w_C^2J_C&\leq& w_A^2J_A+w_B^2J_B\qquad\forall\;w_A,\;w_B\in\mathbb{R}\;,\label{FIn}\\
\quad w_C&=&\sqrt{\lambda_A}\;w_A+\sqrt{\lambda_B}\;w_B\;,
\end{eqnarray}
proven for any beam--splitter (see Methods for the proof in the amplifier case). Our main idea is to choose $w_A$ and $w_B$ in order to get from \eqref{FIn} the strongest possible inequality. For this purpose, we can rewrite $w_C^2$ as
\begin{eqnarray}
w_C^2&=&\left(\sqrt{\frac{\lambda_A}{J_A}}\;w_A\sqrt{J_A}+\sqrt{\frac{\lambda_B}{J_B}}\;w_B\sqrt{J_B}\right)^2\leq\nonumber\\
&\leq&\left(\frac{\lambda_A}{J_A}+\frac{\lambda_B}{J_B}\right)\left(w_A^2J_A+w_B^2J_B\right)\;,
\end{eqnarray}
where we have used the Cauchy--Schwarz inequality. Equality holds iff
\begin{equation}
w_A=k\;\frac{\sqrt{\lambda_A}}{J_A}\;,\qquad w_B=k\;\frac{\sqrt{\lambda_B}}{J_B}\;,\qquad k\in\mathbb{R}\;,
\end{equation}
and with this choice \eqref{FIn} becomes exactly the generalized Stam inequality \eqref{FII}.

Another important and useful property is the quantum analogue of the {\it de Bruijn identity} which relates the Fisher information to the the entropy flow under additive Gaussian noise,
\begin{equation}\label{deB}
 J\equiv \sum_jJ\left(\rho(t)^{(q,P_j)};q\right)+J\left(\rho(t)^{(p,Q_j)};p\right)=4 \frac{d}{dt} S(\rho(t))\,,
 \end{equation}
 where $\rho(t)=e^{\mathcal L t}\rho(0)$ and
 \begin{equation}\label{L}
\mathcal{L}(\rho)\equiv- \frac{1}{4}\sum_{j=1}^n \left([Q_j,[Q_j,\rho]]+[P_j,[P_j,\rho]]\right)\;.
\end{equation}
 The proof, repeated in the Methods, simply follows from the definition of the ensembles  $\rho^{\left(p,Q_j\right)}$ and $\rho^{\left(q,P_j\right)}$ \cite{ks}.

\subsection{Proof of quantum Entropy Power Inequality}
The argument is similar to the one used in the derivation of the classical EPI. This technique, which is based on the addition of white Gaussian noise in the system, was
extended to the quantum domain in Ref.\ \onlinecite{ks} in order to prove the qEPI for the special case of $\lambda=\frac{1}{2}$.
Here we use the properties (\ref{FII}) and (\ref{deB}) of the quantum Fisher information and we show that the qEPI is valid for all $\lambda \in [0,1]$ (beam--splitter) and
for all $\kappa \ge 1$ (amplifier).

The key idea borrowed from the classical proof is to notice that, for highly entropic thermal states, inequalities
 \eqref{qEPI} and  \eqref{qEPIk} are almost saturated.
Then if we evolve the inputs adding classical Gaussian noise, \eqref{qEPI} and \eqref{qEPIk}  will asymptotically hold in the infinite time limit, and we just need to prove that
the added noise has not improved the inequalities.
This can be achieved in the quantum setting by the application of  the Gaussian additive noise channel
\begin{equation}
\rho(t)\equiv e^{t\mathcal{L}}\rho\;,
\end{equation}
where the Liouvillian operator $\mathcal L$ is the one defined in Eq.~(\ref{L}).
We need an asymptotic estimate for the entropy of $\rho(t)$ as $t\to\infty$.
Intuitively, one can guess that for large times the memory of the input state is washed out and that the leading contribution to the entropy comes from the Gaussian noise alone.
Indeed it can be shown (see Methods) that, for every input state $\rho(0)$,
\begin{equation}
e^{S(\rho(t))/n}=\frac{et}{2}+\mathcal{O}(1)\;.\label{asymp}
\end{equation}
We then consider as input states the evolved $\rho_A(t_A)$ and $\rho_B(t_B)$, where we still have the freedom to let $A$ and $B$ evolve with different speeds by suitably choosing the dependence of their times $t_A(t)$ and $t_B(t)$ on a common time $t$, with the conditions:
\begin{eqnarray}
&t_A(0)=t_B(0)=0\;, \\
&t_A,\;t_B\to\infty\quad \text{for}\quad t\to\infty\;.\label{tinf}
\end{eqnarray}
From the composition laws of Gaussian channels, it follows  that evolving $\rho_A$ and $\rho_B$ by times $t_A$ and $t_B$ before the application of the beam--splitter (or of the amplifier)  produces at the output the state $\rho_C$ evolved by a time
\begin{equation}
t_C=\lambda_At_A+\lambda_Bt_B\;.\label{tbs}
\end{equation}
The corresponding time dependent version of the qEPIs (\ref{qEPI}) and (\ref{qEPIk})  can be rearranged in the following form:

\begin{equation}\label{qEPIt}
1\overset{?}{\geq}\frac{\lambda_A e^{S\left[\rho_A(t_A)\right]/n}+\lambda_Be^{S\left[\rho_B(t_B)\right]//n}}{e^{S\left[\rho_C(t_C)\right]/n}}\;.
\end{equation}
Now if we plug in the asymptotic behavior \eqref{asymp}, we see that the inequality is saturated for $t\to \infty$.
The qEPI that we need to prove is simply \eqref{qEPIt} for $t=0$ and this can be achieved if we are able to show that the RHS of (\ref{qEPIt}) is monotonically increasing
in time, {\it i.e.} that
\begin{equation}
\frac{d}{dt}\frac{\lambda_A e^\frac{S_A}{n}+\lambda_Be^\frac{S_B}{n}}{e^\frac{S_C}{n}}\overset{?}{\geq}0\;,\label{ineq}
\end{equation}
where we have put for simplicity
\begin{equation}
S_X=S[\rho_X(t_X)]\quad\text{for}\quad X=A,B,C\;.
\end{equation}
From the {\it quantum de Bruijn identity} (\ref{deB}),
the positivity of the derivative in \eqref{ineq} can be expressed as
\begin{equation}
\lambda_Ae^\frac{S_A}{n}J_A\dot{t}_A+\lambda_Be^\frac{S_B}{n}J_B\dot{t}_B\overset{?}{\geq}\left(\lambda_Ae^\frac{S_A}{n}+\lambda_Be^\frac{S_B}{n}\right)J_C\dot{t}_C\;.\label{ineq2}
\end{equation}
Now we make use of the freedom that we have in choosing the functions $t_A(t)$, $t_B(t)$ and we impose them to satisfy the differential equation
\begin{equation}
\dot{t}_X=e^{S(t_X)/n}\;,\quad X=A,B\; , \label{diff}
\end{equation}
with initial condition
\begin{equation}
t_X(0)=0\;.
\end{equation}
Since the entropy is nonnegative, $\dot{t}_X\geq 1$ and \eqref{tinf} is satisfied.
From Eq.\ (\ref{tbs}) we have  $\dot{t}_C=\lambda_A \dot{t}_A+\lambda_B \dot{t}_B $ and so the condition (\ref{ineq2}) reduces to

\begin{equation}
\left(\lambda_Ae^\frac{S_A}{n}+\lambda_Be^\frac{S_B}{n}\right)^2 J_C\overset{?}{\leq}\lambda_A e^\frac{2S_A}{n} J_A+\lambda_B e^\frac{2S_B}{n} J_B \label{ineq3}\;.
\end{equation}
At this point our quantum version of the {\it Stam inequality} (\ref{FII}) comes into play providing a useful upper bound to $J_C$,
\begin{equation}
J_C\leq\frac{J_AJ_B}{\lambda_AJ_B+\lambda_BJ_A}\;.
\end{equation}
By plugging it into (\ref{ineq3}) and rearranging the terms we get
\begin{equation}
\frac{\lambda_A\lambda_B\left(J_Ae^{S_A/n}-J_Be^{S_B/n}\right)^2}{\lambda_AJ_B+\lambda_BJ_A}\geq0\;,
\end{equation}
which is trivially satisfied because of the non-negativity of the Fisher information.
This concludes the proof of both inequalities (\ref{qEPI}) and (\ref{qEPIk}) and we can now focus on some of
their direct implications.

\subsection{Linear inequalities}
One of the features of the qEPI is that it is a significantly strong bound. For example from the concavity of the logarithm
we directly get from (\ref{qEPI}) and (\ref{qEPIk}) the respective linear relations, \emph{i.e.}
\begin{eqnarray}
S(\rho_C) &\geq & \lambda S(\rho_A)+(1- \lambda)S(\rho_B), \label{blin}\\
S(\rho_C) &\geq&  \frac{\kappa  S(\rho_A) +(\kappa-1)S(\rho_B)}{2\kappa -1}+\ln\left(2 \kappa -1\right). \label{amplin}
\end{eqnarray}
In the classical setting, the analogue of the first of these expressions is known to be formally equivalent to Eq.~(\ref{cEPI}). For the quantum case however, such correspondence is no longer valid and
Eqs.~(\ref{blin})~and (\ref{amplin}) appear to be weaker than~(\ref{qEPI}) and (\ref{qEPIk}), respectively. We remind also that Eq.~(\ref{blin})  was originally conjectured in~\onlinecite{EPnIguha} and proven by K\"{o}nig and Smith in~\onlinecite{ks} for all $\lambda\in [0,1]$.

\paragraph{Bound on the EPnI.--} It turns out that
 Eq.~(\ref{qEPI}) is not the only way of generalizing the classical inequality (\ref{cEPI}). Another possible generalization was proposed and conjectured in Ref. \onlinecite{EPnIguha,guha}. This is the {\it Entropy Photon number Inequality} (EPnI):

\begin{equation} \label{EPnI}
N(\rho_C)  \overset{?}{\ge}  \lambda\; N(\rho_A) + (1- \lambda)\;N(\rho_B)\;,
\end{equation}
where $g(N)=(N+1)\ln(N+1)-N\ln N$ is the entropy of a single mode thermal state with mean photon number $N$, and $N(\rho)= g^{-1}\left(S(\rho)/n\right)$ is the mean photon number per mode of an $n$--mode thermal state with the same entropy of $\rho$. The EPnI states that fixing the input entropies $S_A$, $S_B$, the output entropy $S_C$ is minimum when the inputs are thermal.
Since the qEPI \eqref{qEPI} is \emph{not} saturated by thermal states (unless they have the same entropy), it is weaker than (and it is actually implied by) the EPnI \eqref{EPnI}, so our proof of qEPI does not imply the EPnI, which still remains an open conjecture.
However, as we are going to show, the validity of the qEPI imposes a very tight bound (of the order of $0.132$) on the maximum allowed violation of the EPnI \eqref{EPnI}.

The map  $e^{S(\rho)/n} \mapsto N(\rho)$ from the entropy power to the entropy photon-number is the function $f(x)\equiv g^{-1}(\ln(x))$ defined on the interval $[1,\infty]$. Unfortunately it is convex and we cannot
obtain the EPnI \eqref{EPnI} from~\eqref{qEPI}. Fortunately however, $f(x)$ is {\it not too convex} and is well approximated by a linear function. It is easy to show indeed  that $f(x) =-1/2 + x/e +\delta(x) $, where
$0\leq\delta(x)\le \delta(1)=1/2-1/e \simeq 0.132$. This directly implies that the entropy photon number inequality is valid up to such a small error,
\begin{equation}
N(\rho_C)-\lambda N(\rho_A) - (1- \lambda)N(\rho_B) \ge 1/e -1/2 \;.\label{epnib}
\end{equation}
As a side remark on the EPnI, we conjecture that an inequality similar to \eqref{EPnI} should hold also in the case in
which the mixing channel is the quantum amplifier,
\begin{equation}
N(\rho_C)\overset{?}{\ge}  \kappa\;N(\rho_A) + (\kappa-1)\;\left(N(\rho_B)+1\right)\;,
\end{equation}
but even in this case we do not have a proof.

\subsection{Generalized minimum output entropy conjecture}
\begin{figure}[t]
\includegraphics[width=0.99 \columnwidth]{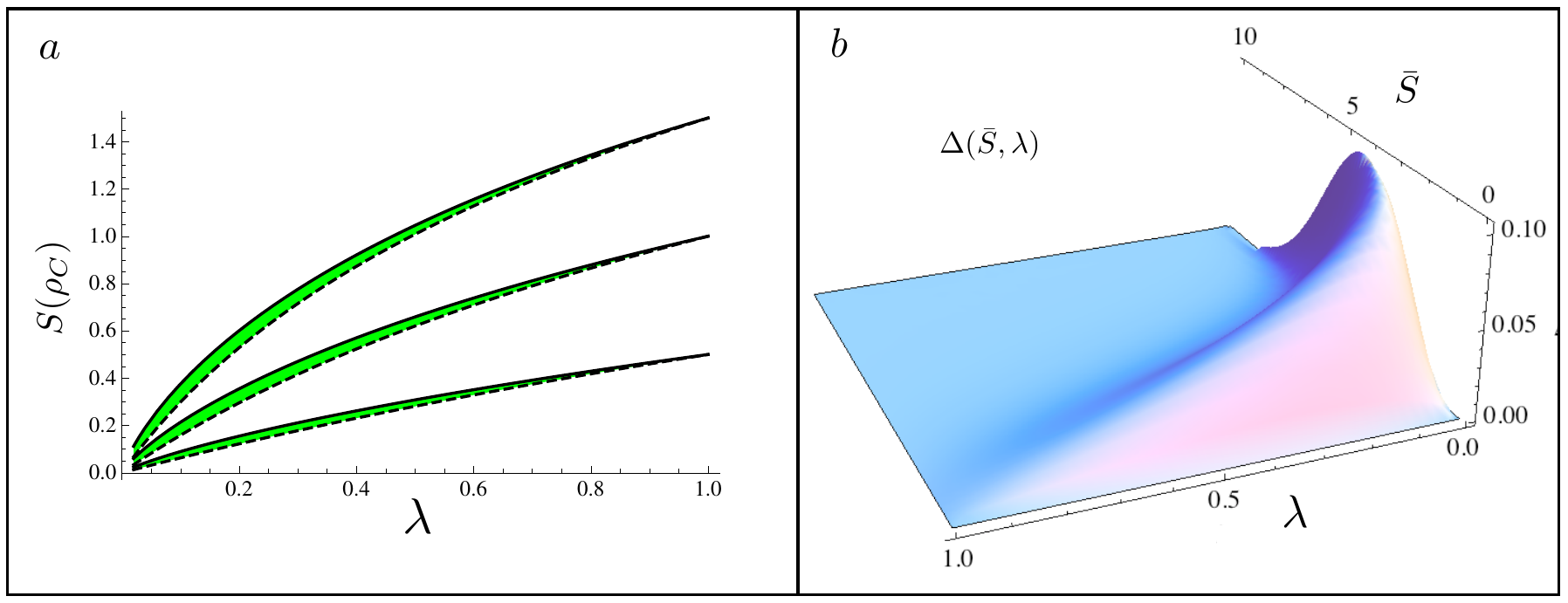}
\caption{{\bf a} Plot of the output entropies as functions of $\lambda$ and for different input entropies $\bar S=0.5,1,1.5$. In full lines are the entropy achievable with a Gaussian input state while the dotted lines represent the lower bound (\ref{bound}). The corresponding minimum output entropies are necessarily constrained within the green regions. Notice that larger values of input entropies $\bar S$ are not considered in this plot because the Gaussian ansatz and the bound becomes practically indistinguishable. {\bf b} Maximum allowed violation $\Delta(\bar S,\lambda)$ of the generalized minimum output entropy conjecture. The two axes are the input entropy  $\bar S$ and the beam--splitter transmissivity $\lambda$. It is evident that the a potential violation of the conjecture is necessarily localized in the parameter space. } \label{delta}
\end{figure}

Recently the so called {\it minimum output entropy conjecture} has been proved \cite{ghgp,mgh,ggpch}. It claims (in the notation of this work) that when $\rho_A$ is a Gaussian thermal state, the minimum output entropy $S(\rho_C)$ is achieved when the input $\rho_B$ is the vacuum. The dual problem \cite{EPnIguha,guha} is to fix $\rho_B=|0\rangle \langle 0 |$ and to ask what is the minimum of $S(\rho_C)$ with the constraint that the input entropy is fixed $S(\rho_A)=\bar{S}>0$. In Ref. \onlinecite{EPnIguha,guha} it was proved that the EPnI \eqref{EPnI} implies that the minimum is achieved by the Gaussian centred thermal state with entropy $\bar{S}$, corresponding to an output entropy of $g\left(\lambda g^{-1}\left(\bar S\right)\right)$. Together with the EPnI, this {\it generalized conjecture} is still an open problem, however we can use our qEPI to obtain a tight lower bound
on $S(\rho_C)$.  The bound follows directly from (\ref{qEPI}) for $S{(\rho_B)}=0$ and can be expressed as
\begin{equation} \label{bound}
S(\rho_C) \ge \ln \left[\lambda\; e^{\bar S}+ (1-\lambda)\right]\;.
\end{equation}
 The RHS of  (\ref{bound}) is extremely close to the conjectured minimum $g\left(\lambda g^{-1}\left(\bar S\right)\right)$. Indeed the error between the two quantities  $\Delta(\bar S,\lambda)=g\left(\lambda  g^{-1}\left(\bar S\right)\right)- \ln \left[\lambda e^{\bar S}+ (1-\lambda)\right]$
 is bounded by $\sim 0.107$ and moreover it decays to zero in large part of the parameter space $(\bar S,\lambda)$ (see Fig.\ \ref{delta}). The plot in Fig.\ \ref{delta} provides also a useful hint
 about the small parameter region where a potential counter-example disproving the conjecture should be looked for.

Our qEPI (\ref{qEPI}), and in particular inequality (\ref{bound}), are also useful for bounding the capacity region of the bosonic broadcast channel. As explicitly discussed in the Methods, this bound is very close to the optimal one~\cite{guhaieee,guhabroadcast}, which however relies on the still unproven conjecture \cite{EPnIguha,guha} mentioned above.

\section{Discussion}

Understanding the complex physics of continuous variable quantum systems~\cite{BRAU} represents a fundamental challenge of modern science which is crucial
for developing an information technology  capable of taking full advantage of quantum effects~\cite{CAVES,WEEDB}.
This task appears now to be within our grasp due to a series of very recent works which have solved a collection  of  long standing conjectures. Specifically,
the minimum output entropy and output majorization conjectures (proposed in Ref.~\onlinecite{CONJ1} and solved in Ref.s~\onlinecite{ghgp} and~\onlinecite{mgh} respectively),
 the optimal Gaussian ensemble and the additivity conjecture (proposed in~\onlinecite{HOWE} and solved in Ref.~\onlinecite{ghgp}),
 the optimality of Gaussian decomposition in the calculation of entanglement of formation~\cite{FORM} and of Guassian discord~\cite{DISC,DISC1}
 for two-mode gaussian states (both solved in Ref.~\onlinecite{ggpch}), the proof of the strong converse of the classical capacity theorem~\cite{STRONG}.
 Our  work represents a fundamental further step in this direction by extending the proof of~\onlinecite{ks} for the qEPI conjecture to include all beam splitter transmissivities and
 by generalizing it to active bosonic transformations (e.g. amplification processes).

\section{Methods}

\subsection{Proof of inequality \texorpdfstring{\eqref{FIn}}{(12)}}
In this section we will prove the inequality \eqref{FIn}
\begin{equation}
w_C^2J_C\leq w_A^2J_A+w_B^2J_B\label{FI}
\end{equation}
for the quantum amplifier. Since the proof is analogue to the beam--splitter case, for clarity we present both.

We start recalling some basic characteristics of these channels.
Their $n$ output modes have annihilation operators
\begin{align}
c_i&=\sqrt{\lambda_A}\; a_i+\sqrt{\lambda_B}\; b_i\quad i=1,\ldots,n\quad\text{(beam--splitter)} \; ,\\
c_i&=\sqrt{\lambda_A}\; a_i+\sqrt{\lambda_B}\; b_i^\dag\quad i=1,\ldots,n\quad\text{(amplifier)}\;,
\end{align}
with $\lambda_A,\;\lambda_B\geq0$ such that
\begin{align}
\lambda_A+\lambda_B&=1\quad\text{(beam--splitter)} \; ,\\
\lambda_A-\lambda_B&=1\quad\text{(amplifier)}\,.
\end{align}
Let $T$ be the time reversal matrix acting on the phase space, which reverses the signs of the quadratures $P_i$ and satisfies
\begin{equation}
T=T^t=T^{-1}\;.
\end{equation}
Let $\gamma_A$ and $\gamma_B$ be the covariance matrices of the two inputs; then the output will have a covariance matrix
\begin{align}
\gamma_C&=\lambda_A\gamma_A+\lambda_B\gamma_B\quad&\text{(beam--splitter)} \; ,\\
\gamma_C&=\lambda_A\gamma_A+\lambda_BT\gamma_BT\quad&\text{(amplifier)}\;.
\end{align}
For the displacement vectors we have instead
\begin{align}
d_C&=\sqrt{\lambda_A}\;d_A+\sqrt{\lambda_B}\;d_B\quad&\text{(beam--splitter)} \; , \label{dispb}\\
d_C&=\sqrt{\lambda_A}\;d_A+\sqrt{\lambda_B}\;Td_B\quad&\text{(amplifier)}\;.\label{dispa}
\end{align}

\subsubsection{Compatibility with the Liouvillian}
We recall \textbf{Lemma III.1} of \onlinecite{ks}: for any $t\geq 0$, the CPTPM $e^{t\mathcal{L}}$ is a Gaussian map acting on covariance matrices and displacement vectors by
\begin{equation}
\begin{matrix}
\gamma&\mapsto \gamma'&=&\gamma+t\mathbbm{1}_{2n} \; ,\\
d&\mapsto d'&=&d \;. \
\end{matrix}
\end{equation}
Then, if we choose as inputs the states $\rho_A(t_A)$ and $\rho_B(t_B)$ evolved with times $t_A$ and $t_B$, the output will have covariance matrix and displacement vector
\begin{align}
\gamma_C(t)&=\lambda_A\gamma_A+\lambda_B\gamma_B+\lambda_At_A\mathbbm{1}_{2n}+\lambda_Bt_B\mathbbm{1}_{2n}=\nonumber\\
&=\gamma_C(0)+t_C\mathbbm{1}_{2n} \; , \\
d_C(t)&=\sqrt{\lambda_A}\;d_A+\sqrt{\lambda_B}\;d_B=d_C(0) \; ,
\end{align}
in the case of the beam--splitter, and
\begin{align}
\gamma_C(t)&=\lambda_A\gamma_A+\lambda_BT\gamma_BT+\lambda_At_A\mathbbm{1}_{2n}+\lambda_Bt_B\mathbbm{1}_{2n}=\nonumber\\
&=\gamma_C(0)+t_C\mathbbm{1}_{2n} \; ,\\
d_C(t)&=\sqrt{\lambda_A}\;d_A+\sqrt{\lambda_B}\;Td_B=d_C(0) \; ,
\end{align}
in the case of the amplifier, where we have used $T^2=\mathbbm{1}_{2n}$ and we have put
\begin{equation}
t_C=\lambda_At_A+\lambda_Bt_B\;.\label{tc}
\end{equation}
Then, first evolving the inputs with times $t_A$, $t_B$ and then applying the beam-splitter / amplifier is the same as first applying the beam--splitter/amplifier and then evolving the output with time $t_C$ as in \eqref{tc}.

\subsubsection{Properties of quantum Fisher information}

To prove inequality \eqref{FI}, we will follow the proof of the beam--splitter version for $\lambda=\frac{1}{2}$ in \onlinecite{ks}.
First, given a smooth parametric family of states $\theta\mapsto \rho^{(\theta)}$, one can define (see Eq.\ (69) in \onlinecite{ks}) the associated quantum Fisher information with
\begin{equation}
J\left(\rho^{(\theta)};\theta\right)\equiv \left.\frac{d^2}{d\theta^2}S\left(\left.\rho^{(0)}\right\|\rho^{(\theta)}\right)\right|_{\theta=0}\;.
\end{equation}
It is linear in the parameter (\onlinecite{ks} \textbf{Lemma IV.1}):
\begin{eqnarray}
J\left(\rho^{(c\theta)};\theta\right)&=c^2 J\left(\rho^{(\theta)};\theta\right)\;,\label{lin}
\end{eqnarray}
and additive on product states (\onlinecite{ks} \textbf{Lemma IV.3}):
\begin{equation}
J\left(\rho_A^{(\theta)}\otimes\rho_B^{(\theta)};\theta\right)=J\left(\rho_A^{(\theta)};\theta\right)+J\left(\rho_B^{(\theta)};\theta\right)\label{add}\;.
\end{equation}
It is also always nonnegative (\onlinecite{ks} \textbf{Lemma IV.2}):
\begin{equation}
J\left(\rho^{(\theta)};\theta\right)\geq0\;,
\end{equation}
and vanishes for $\theta=0$, where it has a minimum. Then the data processing inequality for the relative entropy
\begin{equation}
S(\mathcal{E}(\hat{\rho})\|\mathcal{E}(\hat{\sigma}))\leq S(\hat{\rho}\|\hat{\sigma})
\end{equation}
implies that the quantum Fisher information is non--increasing under the application of any CPTP map $\mathcal{E}$ (\onlinecite{ks} \textbf{Theorem IV.4}):
\begin{equation}
J\left(\mathcal{E}\left(\rho^{(\theta)}\right);\theta\right)\leq J\left(\rho^{(\theta)};\theta\right)\;.\label{dp}
\end{equation}
If the family is generated by conjugation with an exponential as in formula (76) of \onlinecite{ks}:
\begin{equation}
\rho^{(\theta)}=e^{i\theta H}\rho^{(0)}e^{-i\theta H}\;,
\end{equation}
then (\onlinecite{ks} \textbf{Lemma IV.5})
\begin{align}
J\left(\rho^{(\theta)};\theta\right)&=\mathrm{Tr}\left(\rho^{(0)}\left[H,\left[H,\ln\rho^{(0)}\right]\right]\right)=\nonumber\\ &=\mathrm{Tr}\left(\left[H,\left[H,\rho^{(0)}\right]\right]\ln\rho^{(0)}\right)\;.\label{JH}
\end{align}
For $R\in\{Q_j,P_j\}$ we define the displacement operator in the direction $R$ as in \onlinecite{ks}, formula (79):
\begin{eqnarray}
D_R(\theta)=\begin{cases}
e^{i\theta P_j}\qquad &\textrm{ if }R=Q_j \; ,\\
e^{-i\theta Q_j}\qquad &\textrm{ if }R=P_j\ .
\end{cases}
\end{eqnarray}
For a state $\rho$, we consider the family of translated states
\begin{equation}
\rho^{(\theta,\;R)}=D_{R}(\theta)\rho D_{R}(\theta)^\dagger \; ,
\end{equation}
and its Fisher information $J\left(\rho^{(\theta,\;R)};\theta\right)$.
We define the quantity $J(\rho)$ as the sum of the quantum Fisher information along all the phase space directions:
\begin{equation}
J(\rho)\equiv\sum_{k=1}^{2n}J\left(\rho^{(\theta,R_k)};\theta\right)\;.
\end{equation}
Using \eqref{JH}, we get
\begin{align}
J(\rho)= \sum_{i=1}^{n}\mathrm{Tr}\left(\left(\left[P_i,\left[P_i,\rho^{(0)}\right]\right]\right.\right.+\nonumber\\
\left.\left.+\left[Q_i,\left[Q_i,\rho^{(0)}\right]\right]\right)\ln\rho^{(0)}\right)\;,\label{PP}
\end{align}
and since
\begin{equation}
\left.\frac{d}{dt}S\left(e^{t\mathcal{L}}\rho\right)\right|_{t=0}=-\mathrm{Tr}\left(\mathcal{L}(\rho)\ln\rho\right)\;,\label{JL}
\end{equation}
we finally get
\begin{equation}
\frac{dS(\rho(t))}{dt}=\frac{1}{4}J(\rho) \; ,
\end{equation}
as in \textbf{Theorem V.1} of \onlinecite{ks}.
The key point here is that if we define $\widetilde{J}$ with the time inverted quadratures
\begin{equation}
\widetilde{J}(\rho)\equiv\sum_{k=1}^{2n}J\left(\rho^{(\theta,TR_k)};\theta\right)\;,
\end{equation}
the two definitions coincide:
\begin{equation}
\widetilde{J}(\rho)=J(\rho)\label{reveq} \; ,
\end{equation}
since the $P_j$ appear always quadratically in \eqref{PP}.

We now want to apply the data processing inequality \eqref{dp} to our beam--splitter/amplifier channel to obtain the quantum Fisher information inequality.

\subsubsection{Compatibility with translations}
Let $\mathcal{E}$ be the channel associated with the beam--splitter/amplifier.
Then
\begin{equation}
\mathcal{E}\left(\rho_A^{(w_A\theta,R)}\otimes\rho_B^{(w_B\theta,R)}\right)=\mathcal{E}\left(\rho_A\otimes\rho_B\right)^{(w_C\theta,R)}\; ,   \label{trcompb}
\end{equation}
for the beam--splitter, and
\begin{equation}\label{trcompa}
\mathcal{E}\left(\rho_A^{(w_A\theta,R)}\otimes\rho_B^{(w_B\theta,TR)}\right)=\mathcal{E}\left(\rho_A\otimes\rho_B\right)^{(w_C\theta,R)} \; ,
\end{equation}
for the amplifier, \emph{i.e.} translating the inputs by
\begin{align}
d_A=w_A\theta d_R & \quad d_B=w_B\theta d_R & \quad\text{(beam--splitter)} \; ,\\
d_A=w_A\theta d_R & \quad d_B=w_B\theta Td_R & \quad\text{(amplifier)} \; ,
\end{align}
(notice the time reversal) and then applying the beam--splitter/amplifier is the same as applying the beam--splitter/amplifier and translating the output by $w_C\theta d_R$, where $d_R$ is the phase space vector associated to the operator $R$ and
\begin{equation}
w_C=\sqrt{\lambda_A}\;w_A+\sqrt{\lambda_B}\;w_B\;.
\end{equation}

The proof follows straightforwardly evaluating the displacement vectors with \eqref{dispb}, \eqref{dispa}: if we translate the inputs and then apply the beam--splitter we have
\begin{align}
d_A(\theta)&=d_A+w_A\theta d_R\; , \\
d_B(\theta)&=d_B+w_B\theta d_R\; , \\
d_C(\theta)&=\sqrt{\lambda_A}\;d_A(\theta)+\sqrt{\lambda_B}\;d_B(\theta)=\nonumber\\
&=\sqrt{\lambda_A}\;d_A+\sqrt{\lambda_B}\;d_B+\sqrt{\lambda_A}\;w_A\theta d_R+\nonumber\\
&\qquad{}+\sqrt{\lambda_B}\;w_B\theta d_R=\nonumber\\
&=d_C(0)+w_C\theta d_R\;,
\end{align}
which is what we would get translating the output by $w_C\theta d_R$. The same happens for the amplifier:
\begin{align}
d_A(\theta)&=d_A+w_A\theta d_R \; ,\\
d_B(\theta)&=d_B+w_B\theta Td_R \; , \\
d_C(\theta)&=\sqrt{\lambda_A}\;d_A(\theta)+\sqrt{\lambda_B}\;Td_B(\theta)=\nonumber\\
&=\sqrt{\lambda_A}\;d_A+\sqrt{\lambda_B}\;Td_B+\sqrt{\lambda_A}\;w_A\theta d_R+\nonumber\\
&\qquad{}+\sqrt{\lambda_B}\;w_B\theta d_R=\nonumber\\
&=d_C(0)+w_C\theta d_R\;.
\end{align}
Now we can apply the data processing inequality \eqref{dp} to \eqref{trcompb} and \eqref{trcompa}. Using the additivity \eqref{add} and the linearity \eqref{lin} of the Fisher information, we get
\begin{equation}
w_C^2J\left(\rho_C^{(\theta,R)};\theta\right)\leq w_A^2J\left(\rho_A^{(\theta,R)};\theta\right)+w_B^2J\left(\rho^{(\theta,R)};\theta\right) \; ,
\end{equation}
for the beam--splitter, and
\begin{equation}
w_C^2J\left(\rho_C^{(\theta,R)};\theta\right)\leq w_A^2J\left(\rho_A^{(\theta,R)};\theta\right)+w_B^2J\left(\rho^{(\theta,TR)};\theta\right) \; ,
\end{equation}
for the amplifier. These two results are identical, apart from the time reversal in $B$ in the amplifier case.
Finally, summing over the phase space direction we get in both cases the desired inequality
\begin{equation}
w_C^2J_C\leq w_A^2J_A+w_B^2J_B\;,
\end{equation}
since we have proved in \eqref{reveq} that the time reversal does not affect the sum.

\subsection{Proof of the asymptotic scaling \texorpdfstring{\eqref{asymp}}{(19)}}
In \onlinecite{ks}, {\bf{Corollary III-4}} it is shown that
\begin{equation}
\exp\left(\frac{1}{n}S\left(e^{t\mathcal{L}}\hat{\rho}\right)\right)\geq \frac{et}{2}+\mathcal{O}(1)\;;
\end{equation}
here we prove the upper bound.

Let $\hat{\rho}_G$ be the Gaussianized version of $\hat{\rho}$, \emph{i.e.} the Gaussian state with the same first and second moments. Since Gaussianization always increases entropy \cite{gauss} and commutes with the Liouvillean $\mathcal{L}$ \cite{ks}, $S\left(e^{t\mathcal{L}}\hat{\rho}\right)\leq S\left(e^{t\mathcal{L}}\hat{\rho}_G\right)$.
The covariance matrix of $e^{t\mathcal{L}}\hat{\rho}$ and $e^{t\mathcal{L}}\hat{\rho}_G$ is (\onlinecite{ks}, {\bf{Lemma III.1}}) $\sigma+t\mathbbm{1}_{2n}$, where $\sigma$ is the one of $\hat{\rho}$.
Let $\lambda_0$ be the maximum eigenvalue of $\sigma$. Then $\sigma+t\mathbbm{1}_{2n}\leq(\lambda_0+t)\mathbbm{1}_{2n}$, i.e. the Gaussian thermal state with covariance matrix $(\lambda_0+t)\mathbbm{1}_{2n}$ can be obtained adding (non--white) Gaussian noise to $e^{t\mathcal{L}}\hat{\rho}_G$. Since the additive noise channel is unital, it always increases the entropy, and
\begin{equation}
S\left(e^{t\mathcal{L}}\hat{\rho}_G\right)\leq n g\left(\frac{\lambda_0+t-1}{2}\right)\;.
\end{equation}
Since
\begin{equation}
g\left(x-\frac{1}{2}\right)=\ln(ex)+\mathcal{O}\left(\frac{1}{x^2}\right)\;\text{for}\;x\to\infty\;,
\end{equation}
putting all together we get
\begin{equation}
\exp\left(\frac{1}{n}S\left(e^{t\mathcal{L}}\hat{\rho}\right)\right)\leq \frac{et}{2}+\mathcal{O}(1)\;.
\end{equation}

\subsection{Capacity region of the bosonic broadcast channel}
\begin{figure}[ht]
\includegraphics[width=0.6\textwidth]{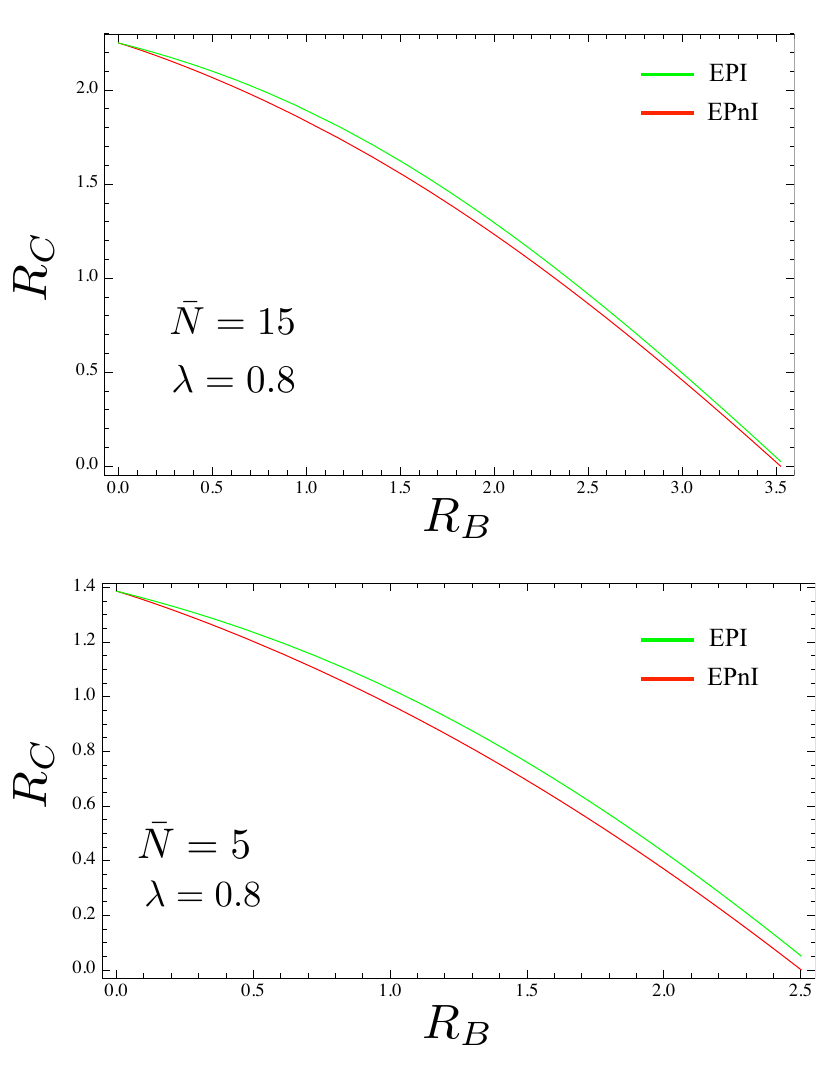}
\caption{Capacity region (expressed in nats per channel uses) for a broadcasting channel~\cite{guhaieee,guhabroadcast} in which the sender is communicating simultaneously with two receivers ($B$ and $C$) via a single bosonic mode which
splits at a beam splitter of transmissivity $\lambda$  ($B$ receiving the transmitted signals, while $C$ receiving the reflected one), under input energy constraint which limits the
mean photon number of the input messages to be smaller than $\bar{N}$. The region delimited by the red curve represents the achievable rates $R_B$ and $R_C$ which would apply if the (still unproven) EPnI conjecture~\eqref{EPnI} held. The green curve instead is the bound one can derive via Eq.~(\ref{bound}) from the EPI inequality~(\ref{qEPI}) we have proven in this paper.}
 \label{fig1}
\end{figure}

In Ref.~\onlinecite{guhaieee,guhabroadcast} it is proven that, trusting the minimum output entropy conjecture of \onlinecite{EPnIguha,guha} (which is a particular case of the still unproven EPnI), the capacity region for a lossless bosonic broadcast channel is parametrically described by the inequalities
\begin{eqnarray}
R_B&\leq& g\left(\lambda\beta\bar{N}\right)\;,\label{rfinale}
  \\ \nonumber
R_C&\leq& g\left((1-\lambda)\bar{N}\right)-g\left((1-\lambda)\beta\bar{N}\right)\;, \end{eqnarray}
with $R_B$ and $R_C$ representing the achievable communication rates the sender of the information can establish when signaling simultaneously to
two independent receivers $B$ and $C$, respectively, when coding
 his messages into a single bosonic mode which splits at a beam splitter of  transmissivity $\lambda\geq\frac{1}{2}$  (the transmitted signals being routed to $B$ and the reflected ones to $C$, see~\onlinecite{guhaieee,guhabroadcast} for details).
In this expression  $\beta \in [0,\,1]$ represents the fraction of the sender's average photon number that is meant to convey information to $B$, with the remainder to be used to communicate information to $C$.  $\bar{N}\geq 0$ instead  is the maximum average mean input photon number employed in the communication per channel uses.
Our qEPI inequality~(\ref{qEPI}) provides instead the weaker bound
\begin{eqnarray}
R_B&\leq& g\left(\lambda\beta\bar{N}\right)\;, \label{rfinale2} \\
R_C&\leq& g\left((1-\lambda)\bar{N}\right)-\ln\frac{(1-\lambda)e^{g\left(\lambda\beta\bar{N}\right)}+2\lambda-1}{\lambda}\;.\nonumber
\end{eqnarray}
A comparison between Eq.~(\ref{rfinale2}) and the conjectured region~(\ref{rfinale})  is shown in  Fig.~\ref{fig1}: the discrepancy  being small.

\section{Acknowledgements}
The authors are grateful to L. Ambrosio, S. Di Marino, and A. S. Holevo for comments and
discussions. AM acknowledges support from Progetto Giovani Ricercatori 2013 of Scuola Normale Superiore.

\section{Supplementary Material}

\subsection{Linear inequality for the quantum amplifier}
The proof of the inequality \eqref{amplin}
\begin{equation}
S(\rho_C) \geq  \frac{\kappa  S(\rho_A) +(\kappa-1)S(\rho_B)}{2\kappa -1}+\ln\left(2 \kappa -1\right) \; ,
\end{equation}
for the amplifier is straightforward: the entropy power inequality \eqref{qEPIk} can be rewritten as
\begin{equation}
S_C\geq\ln\left(\frac{\lambda_A}{\lambda_A+\lambda_B}e^{S_A}+\frac{\lambda_B}{\lambda_A+\lambda_B}e^{S_B}\right)+\ln\left(\lambda_A+\lambda_B\right)\,,
\end{equation}
which for the concavity of the logarithm implies
\begin{equation}
S_C\geq\frac{\lambda_AS_A+\lambda_BS_B}{\lambda_A+\lambda_B}+\ln\left(\lambda_A+\lambda_B\right)\;,
\end{equation}
\emph{i.e.} \eqref{amplin}.

This result can also be proven without recurring to the qEPI: let us evolve with the Liouvillian the inputs with equal times
\begin{equation}
t_A=t_B=t\;.
\end{equation}
The corresponding evolution time for the output will be
\begin{equation}
t_C=(\lambda_A+\lambda_B)t\;.
\end{equation}
Recalling the asymptotic behaviour of the entropies \eqref{asymp}, both sides of \eqref{amplin} behave as
\begin{equation}
\ln\frac{et}{2}+\ln\left(\lambda_A+\lambda_B\right)+\mathcal{O}\left(\frac{1}{t}\right)
\end{equation}
for $t\to\infty$, and \eqref{amplin} is asymptotically saturated.
Then we have only to check that
\begin{equation}
\frac{d}{dt}S_C\leq\frac{d}{dt}\frac{\lambda_AS_A+\lambda_BS_B}{\lambda_A+\lambda_B}\;,
\end{equation}
\emph{i.e.} that
\begin{equation}
J_C\leq\frac{\lambda_AJ_A+\lambda_BJ_B}{(\lambda_A+\lambda_B)^2}\;,\label{ineq4}
\end{equation}
where we have used
\begin{equation}
\frac{d}{dt}S_C=(\lambda_A+\lambda_B)J_C\;.
\end{equation}
But \eqref{ineq4} is exactly what we get if we plug in the quantum Fisher information inequality \eqref{FI}
\begin{align}
w_A&=\sqrt{\lambda_A} \; ,\\
w_B&=\sqrt{\lambda_B}\;.
\end{align}

\subsection{Bound on EPnI}
We want to evaluate how close is qEPI \eqref{qEPI} to EPnI \eqref{EPnI} and prove \eqref{epnib}.
\eqref{qEPI} implies for the output entropy photon number
\begin{equation}
N_C\geq g^{-1}\left(\ln\left(\lambda_A e^{g(N_A)}+\lambda_Be^{g(N_B)}\right)\right)\label{nc}\;.
\end{equation}
\eqref{EPnI} is stronger than \eqref{qEPI}, and in fact
\begin{equation}
g^{-1}\left(\ln\left(\lambda_A e^{g(N_A)}+\lambda_Be^{g(N_B)}\right)\right)\leq \lambda_AN_A+\lambda_BN_B\;,
\end{equation}
since the function $g^{-1}\left(\ln\left(x\right)\right)$ is increasing and convex.
Since $e^{g(N)}$ for $N\to\infty$ goes like
\begin{equation}
e^{g(N)}=e\left(N+\frac{1}{2}\right)+\mathcal{O}\left(\frac{1}{N}\right)\;,
\end{equation}
we have for $x\to\infty$
\begin{equation}
g^{-1}\left(\ln x\right)=\frac{x}{e}-\frac{1}{2}+\mathcal{O}\left(\frac{1}{x}\right)\;.
\end{equation}
If we define as in the main text
\begin{equation}
\delta(x)\equiv g^{-1}\left(\ln x\right)-\frac{x}{e}+\frac{1}{2}\;,
\end{equation}
$\delta$ is convex, decreasing and
\begin{equation}
\lim_{x\to\infty}\delta(x)=0\;.
\end{equation}
We can also evaluate
\begin{equation}
\delta(1)=\frac{1}{2}-\frac{1}{e}\;,
\end{equation}
and for any $x_A,\;x_B\geq1$ we have
\begin{equation}
\delta(\lambda_Ax_A+\lambda_Bx_B)\geq\lambda_A\delta(x_A)+\lambda_B\delta(x_B)-\left(\frac{1}{2}-\frac{1}{e}\right)\;.
\end{equation}
Since
\begin{eqnarray}
&g^{-1}\left(\ln\left(\lambda_Ax_A+\lambda_Bx_B\right)\right)-\nonumber\\
&-\lambda_Ag^{-1}\left(\ln x_A\right)-\lambda_Bg^{-1}\left(\ln x_B\right)=\nonumber\\
&=\delta\left(\lambda_Ax_A+\lambda_Bx_B\right)-\lambda_A\delta(x_A)-\lambda_B\delta(x_B)\;,
\end{eqnarray}
in the case $x_A=e^{S_A}$, $x_B=e^{S_B}$ we get
\begin{eqnarray}
g^{-1}\left(\ln\left(\lambda_Ae^{S_A}+\lambda_Be^{S_B}\right)\right)-\lambda_AN_A-\lambda_BN_B=\nonumber\\
=\delta\left(\lambda_Ae^{S_A}+\lambda_Be^{S_B}\right)-\lambda_A\delta(e^{S_A})-\lambda_B\delta(e^{S_B})\;,
\end{eqnarray}
and we can conclude from \eqref{nc} that
\begin{align}
N_C & \geq \lambda_AN_A+\lambda_BN_B+\nonumber\\
&\quad{}+\delta(\lambda_Ae^{S_A}+\lambda_Be^{S_B})-\lambda_A\delta(e^{S_A})-\lambda_B\delta(e^{S_B})\geq\nonumber\\
&\geq\lambda_AN_A+\lambda_BN_B-\left(\frac{1}{2}-\frac{1}{e}\right)\;,
\end{align}
so the \eqref{EPnI} violation can be at most
\begin{equation}
\frac{1}{2}-\frac{1}{e}\simeq 0.132\;.
\end{equation}

\begin{thebibliography}{99}


\bibitem{Shannon}  Shannon C. E., A Mathematical Theory of Communication, Bell Syst. Tech. J. 27, 379-423 (1948).

\bibitem{dembo} Dembo, A., Cover, T. \& Thomas, J., Information theoretic inequalities. IEEE Trans. Inf. Theory 37, 1501-1518 (1991).

\bibitem{stam} Stam, A., Some inequalities satisfied by the quantities of information of Fisher and Shannon. Inform. Control 2, 101-112 (1959).

\bibitem{blachman} Blachman, N., The convolution inequality for entropy powers. IEEE Trans. Inf. Theory 11, 267-271 (1965).

\bibitem{verdu} Verd\'{u}, S. \& Guo, D., A simple proof of the entropy-power inequality. IEEE Trans. Inf. Theory 52, 2165-2166 (2006).

\bibitem{rioul} Rioul, O., Information theoretic proofs of entropy power inequalities. IEEE Trans. Inf. Theory 57, 33-55 (2011).

\bibitem{guo} Guo, D., Shamai, S. \& Verd\'{u}, S. Proof of Entropy Power Inequalities Via MMSE in Proceedings of Information Theory 2006, IEEE International Symposium on Information Theory, 1011-1015 (2006).

\bibitem{bergmans} Bergmans, P., A simple converse for broadcast channels with additive white Gaussian noise (corresp.). IEEE Trans. Inf. Theory 20, 279-280 (1974).

\bibitem{LYC} Leung-Yan-Cheong, S. \& Hellman, M., The Gaussian wire-tap channel. IEEE Trans. Inf. Theory 24, 451-456 (1978).

\bibitem{EPnIguha} Guha, S., Erkmen, B.I. \& Shapiro, J.H., The Entropy Photon-Number Inequality and its consequences. Information Theory and Applications Workshop, 128-130 (2008)

\bibitem{guha} Guha, S., Shapiro, J. \& Erkmen, B., Capacity of the bosonic wiretap channel and the entropy photon-number inequality in Proceedings of Information Theory 2008, IEEE International Symposium on Information Theory, 91-95 (2008).

\bibitem{guhaieee} Guha, S. \& Shapiro, J., Classical Information Capacity of the Bosonic Broadcast Channel in Proceedings of Information Theory 2007, IEEE International Symposium on Information Theory, 1896-1900 (2007).

\bibitem{guhabroadcast} Guha S., Shapiro, J. H. \& Erkmen, B. I., Classical capacity of bosonic broadcast communication and a minimum output entropy conjecture. Phys. Rev. A 76,
032303 (2007).


\bibitem{ghgp}
  Giovannetti V., Holevo, A. S. \& Garc{\'i}a-Patr\'on, R.,
  A solution of the Gaussian optimizer conjecture. Preprint at http://lanl.arxiv.org/abs/1312.2251 (2013).

\bibitem{ggpch}
  Giovannetti V., Garc{\'i}a-Patr\'on R., Cerf, N. J. \& Holevo A. S.,
  Ultimate communication capacity of quantum optical channels by solving the Gaussian minimum-entropy conjecture. Preprint at http://lanl.arxiv.org/abs/1312.6225 (2013).

\bibitem{CONJ1}
Giovannetti V., Guha, S., Lloyd, S., Maccone, L. \& Shapiro, J. H.,
Minimum output entropy of bosonic channels: A conjecture. Phys. Rev. A 70,  032315 (2004).


\bibitem{ks} K\"onig R. \& Smith G., The Entropy Power Inequality for Quantum Systems. IEEE Trans. Inf. Theory 60, 1536-1548 (2014).


\bibitem{ks-natphot}
K\"onig R. and Smith G., Limits on classical communication from quantum entropy power inequalities. Nature Photon. 7, 142-146 (2013).



\bibitem{CLT} Barron, A. R., Entropy and the Central Limit Theorem.
Ann. Prob. 14, 336-342 (1986).



\bibitem{BRAU}
Braunstein S. L. and van Loock P.,
Quantum information with continuous variables.
Rev. Mod. Phys. 77, 513-577 (2005).


\bibitem{WEEDB}  Weedbroock C. et al.,
Gaussian Quantum Information. Rev. Mod. Phys. 84, 621-669 (2012).


\bibitem{WALLS} Walls D. F. \& Milburn G. J., Quantum Optics (Springer, 1994).



\bibitem{channels} Holevo, A. S., Quantum systems, channels, information: a
mathematical introduction, (Walter de Gruyter, 2012).


\bibitem{STAMINEQ} Kagan, A. \& Yu, T., Some inequalities related to the Stam inequality. Appl. Math. 53, 195-205 (2008).




 \bibitem{mgh}
  Mari, A., Giovannetti V. \& Holevo, A.~S.,
  Quantum state majorization at the output of bosonic Gaussian channels.
  Nature Commun. 5, 3826 (2014).






\bibitem{CAVES}
Caves C. M. \& Drummond P. B.,
Quantum limits on bosonic communication rates,
Rev. Mod. Phys. 66,  481-537 (1994).





\bibitem{HOWE} Holevo A. S. \& Werner R. F.,
Evaluating capacities of bosonic Gaus- sian channels, Phys. Rev. A 63,  032312 (2001).


\bibitem{FORM}
Giedke G., Wolf, M. M., Kr\"{u}ger, O.,  Werner, R. F. \& Cirac, J. I., Entanglement of Formation for Symmetric Gaussian States, Phys. Rev. Lett. 91, 107901 (2003).


\bibitem{DISC}
Modi, K., Brodutch, A., Cable, H., Paterek, T. \& Vedral, V., The classical-quantum boundary for correlations: Discord and related measures. Rev. Mod. Phys. 84, 1655-1707 (2012).


\bibitem{DISC1}
Pirandola, S., Cerf, N. J., Braunstein, S. L. \& Lloyd, S., Optimality of Gaussian discord. Preprint at http://lanl.arxiv.org/abs/1309.2215 (2013).


\bibitem{STRONG}
Bardhan, B. R., Garc{\'i}a-Patr\'on, R., Wilde, M. M. \& Winter, A.,
Strong converse for the classical capacity of optical quantum communication channels. Preprint at http://lanl.arxiv.org/abs/1401.4161 (2014).

\bibitem{gauss}
M. M. Wolf, G. Giedke, \& J. I. Cirac, Extremality of Gaussian
quantum states, Phys. Rev. Lett. 96, 080502 (2006).

\end{thebibliography}
\end{document}